\documentclass[aps,pre,reprint,groupedaddress]{revtex4-1}
\usepackage{graphicx}

\begin{document}


\title{A path integral approach for the colloidal glass transition based on an analogy with the $\lambda$-transition in liquid helium}


\author{V. Prasad}
\affiliation{Core R\textsl{\&}D, The Dow Chemical Company, Midland, MI 48674}


\date{\today}

\begin{abstract}
We describe a model for the colloidal glass transition that is based on Feynman's theory for the $\lambda$-transition in liquid helium. Our model essentially counts the number of configurations of dynamic loops, strings or clusters of different sizes, and determines the glass transition volume fraction $\phi_{g}$ from the distribution of these heterogeneities. Since confinement restricts the available number of configurations for these loops, strings and clusters, its effect on $\phi_{g}$ can also be calculated in a relatively straightforward manner .
\end{abstract}


\maketitle

\section{Introduction}
In 1953, Feynman \cite{feyn53, feyn88} proposed a model for the $\lambda$-transition in liquid helium, based on the approach of writing the partition function as a path integral over trajectories of atoms. In this approach, it is assumed that the motion of one atom is not opposed by a potential barrier, but rather facilitated by other atoms moving out of the way. Because of this, a simpler form for the partition function can be written (independent of the interaction potential), which then shows a transition because of the symmetry statistics of the helium atoms. Essentially, atoms can be permuted with each other where these permutations take the form of loops, with the loops having different perimeters (with different number of atoms $s$, $s$ going from 1 to $\infty$). Of all the loops with a given $s$, the dominant contribution comes from a loop where the atoms are nearest neighbors. We will use this concept, and the form of the partition function proposed for liquid helium, to describe a transition that shows remarkable similarities to the colloidal glass transition. In order to do so, we will have to describe the Feynman model in some detail and argue by analogy.

The partition function for liquid helium, obtained by the path integral method (sum over all paths of $e^{iS/\hbar}$, where $S$ is the action) and assuming symmetry statistics is exactly \cite{feyn53}
\begin{eqnarray}\label{helium}
Q = \frac{1}{N!}\sum_{P}\int d^{N}z_{i} \int_{tr P}
e^{-\int_{0}^{\beta} [\frac{m}{2\hbar^{2}}\sum_{i}(\frac{dx_{i}}{du})^{2}} \nonumber\\
^{+\sum_{ij}V(x_{i}-x_{j})]du}\mathcal{D}^{N}x_{i}(u)
\end{eqnarray}

In this partition function, the initial co-ordinates of helium atoms are assumed to be $x_{i}(0) = z_{i}$, but the final co-ordinates are not, instead being a permutation of these denoted by $Pz_{i}$. The integral $\int_{tr P}$ is taken over all trajectories $x_{i}(u)$ of the particles such that $x_{i}(0)=z_{i}$, $ x_{i}(\beta)=Pz_{i}$ . The sum is taken over all permutations $P$ and the integral over all configurations $z_{i}$. Further noted by Feynman was that the quantity $\beta$ ($=it/\hbar$) was not the time, but a vivid representation of Eq.~\ref{helium} could be obtained by assuming that it was the time. Within this framework, one can note that the particles at time 0 form an initial configuration $z_{i}$, and the particles move about such that at time $\beta$ their configuration is nearly the same, except that some of the particles may have been interchanged.

Feynman then postulated that the motion of a particle, even in the presence of strong interparticle potentials, can be described as that of a free particle. This motion involves rearrangement of the positions of other particles in its vicinity to make room for it. Then, for every trajectory, a helium atom behaves as a free particle with effective mass $m'$, where $m'$ incorporates the effects of the rearrangements of the other particles. Since the interparticle potential $V\simeq0$, the time integral over all paths $x_{i}$ for atom $i$ to go a certain distance $a$ is proportional to $(m'/2\pi\beta\hbar^{2})^{3/2}e^{-m'a^{2}/2\beta\hbar^{2}}$. As the final position of particle $x_{i}(\beta)=Pz_{i}$, the partition function in Eq.~\ref{helium} can be approximated as~\cite{feyn53}
\begin{eqnarray}\label{helium2}
Q=\frac{K_{\beta}}{N!}(\frac{m'}{2\pi\beta\hbar^{2}})^{3N/2}\int\sum_{P} e^{[-\frac{m'}{2\beta\hbar^{2}}\sum_{i}(z_{i}-Pz_{i})^2]} \nonumber\\
\times\rho(z_{1}, z_{2},...z_{N})d^{N}z_{i}
\end{eqnarray}
where $K_{\beta}$ is a normalization constant and the function $\rho(z_{1},...z_{N})$ is a configurational density.

Equation~\ref{helium2} is a partition function based on quantum mechanical principles and the notion of imaginary time. How then can we relate this back to non-quantum mechanical systems such as colloids? First, consider a transformation where we replace $it$ by $\tau$ and $\hbar/m'$ with a diffusion constant $D$ (and recall that $\beta=it/\hbar$). Based on such a double transformation, we can rewrite $m'/(2\beta\hbar^{2})=m'/(2 it\hbar)=1/(2D\tau)$. Second, unlike colloids, quantum mechanical particles can permute with each other regardless of any geometrical constraints, such as the distance between these particles. However, each permutation may be divided into loops, where a loop of length $s$ is a chain of permutations, such that particle 1 goes to 2, 2 goes to 3, 3 goes to 4, etc. until finally particle $s$ goes to 1. Further, the dominant contribution to Eq.~\ref{helium2} near the $\lambda$-transition occurs when the particles in loops are nearest neighbours to each other. For colloids, instead of permutations ($Pz_{i}$), consider the classical case of \textbf{\emph{rearrangements}} ($Rz_{i}$) where a particle simply displaces one of its nearest neighbors, which then displaces a third particle and so on. The partition function in Eq.~\ref{helium2} can now be rewritten for the specific colloidal case of rearrangements of particles with their neighbors as
\begin{eqnarray}\label{colloid}
Q=\frac{K_{\tau}}{N!}(\frac{1}{2\pi D\tau})^{3N/2}\int\sum_{R}e^{[-\frac{1}{2D\tau}\sum_{i}(z_{i}-Rz_{i})^{2}]} \nonumber \\
\times\rho(z_{1}, z_{2},...z_{N})d^{N}z_{i}
\end{eqnarray}
where $K_{\tau}$ is a normalization constant, and the sum $R$ is over all rearrangement configurations. Note that the exponential term in the integral for a given $i$ is the transition probability for a colloid to diffuse the distance to its nearest neighbor. Therefore, an alternate, more intuitive way to describe Eq.~\ref{colloid} is simply the transition probability for all particles to diffuse the distance to their neighbors simultaneously, summed over all configurations and weighed by each configurational density. Equation ~\ref{colloid} is of great generality, and can be applied towards situations where colloids form dynamic loops, strings and even clusters.

For the analogy between Eq.~\ref{helium2} and Eq.~\ref{colloid} to work, it is critical to demonstrate a one-to-one correspondence between the liquid helium and colloidal cases. In the Feynman model, the `time' $\beta$ was chosen such that all particles had ended in their final configurations as permutations of their initial configurations. For the colloidal case, $\tau$ is the time for rearrangements of particle positions with their neighbors to occur \cite{Glotzer98, Weeks00, Berthier11}. Our approximate dynamic partition function will therefore be most accurate for this time slice $\tau$ and less accurate at other times, where the numbers of loops or strings will not be at a maximum. The diffusion constant $D$ will be smaller than the free diffusion constant $D_{0}$, just as $m'$ is larger than $m$ in the liquid helium case, because other particles have to move out of the way. Lastly, note the presence of the factor $N!$ in the denominator of Eq.~\ref{colloid}. For a discussion of why this is correct for distinguishable particles such as colloids see references \cite{Swendsen, Frenkel, Cates}.

If the partition function in Eq.~\ref{helium2} demonstrates a transition (as shown by Feynman and others), then it stands to reason that the `dynamic' partition function in Eq.~\ref{colloid} must also demonstrate a transition. While the former is a transition in temperature, because of the relation $\beta=it/\hbar$ and the Wick rotation $it\rightarrow \tau$, it should be possible to relate it to a dynamic transition (i.e. in time $\tau$). The `strings' of particle motions that are required for Eq.~\ref{colloid} have been seen near the glass transition in atomic/molecular systems \cite{Glotzer98}, and granular materials \cite{Durian07}, although not exactly in colloidal systems where dynamic clusters rather than strings are predominant \cite{Weeks00, Zhu12}. Nevertheless, we will build our path integral theory by considering various cases, beginning with loops, then strings, and finally clusters. Further, in all experiments and simulations to date on molecular and colloidal systems \cite{Glotzer98, Weeks00}, the number of particles participating in correlated, string-like motion is only a fraction $f$ of the total number of particles. Again, we will begin by assuming \textbf{\emph{all}} particles participate in loop, string or cluster-like motions. We will then later show this makes no substantial difference to the existence of a transition. That is, if instead of $N$ particles, only a fraction $fN$ of them participate in loops, strings or clusters, then the nature of the transition will remain the same. Finally, since our approach relies on the counting of configurations, it should be relatively straightforward to extrapolate it to the case of confinement, where the sample size is finite, say a box of dimension $L$. We will also provide some preliminary calculations relating to how the glass transition is affected by such a confinement. In subsequent sections, we will calculate the colloidal glass transition volume fraction $\phi_{g}$ for the various situations described above.

\section{Theory}
\subsection{Loop glass transition, all particles participating}
A calculation of $\phi_{g}$ for this situation is the most straightforward, if we start with the liquid helium case and closely follow Feynman's approach \cite{feyn53, feyn88}. In Eq.~\ref{helium2}, consider only permutations involving shifts of a particle to its neighbor. The exponential factor from such a shift would be $y=e^{-m'd^{2}/2\beta\hbar^{2}}$, where $d$ is the mean interparticle spacing. For the colloidal case in Eq.~\ref{colloid}, the corresponding exponential factor will instead be $y=e^{-d^2/2D\tau}$. Further assume that the function $\rho$ is 0 when particles are less than a distance $d$ apart, and 1 everywhere else. Each permutation can be broken up into loops, where a loop of $s$ particles will have a contribution of $y^{s}$. If we have $n_{2}$ loops of 2 particles, $n_{3}$ loops of 3 particles and so on, the contribution to the partition function is $y^{2n_{2}+3n{3}+...}$. The transition determining part of the partition function is therefore
\begin{equation}\label{transition1}
q=\sum G(n_{2},n_{3},...)y^{\sum_{s=2} sn_{s}}
\end{equation}
where $G(n_{2},n_{3},...)$ is the number of permutations with $n_{s}$ loops with $s$ particles.
This sum is difficult to evaluate because of the constraint $\sum sn_{s}=N$. Since the loops compete for the available particles, we can assume that there is an average probability $p$ that each site is unoccupied. Therefore, instead of $q$, we calculate $q'=p^{N}q=e^{-B/kT}$, eliminating this constraint. If $R_{s}$ is the total number of loops of size $s$ that can made, then each loop can be chosen in $R_{s}$ ways, and all $n_{s}$ of them in $R_{s}^{n_{s}}/n_{s}!$ ways (that is, $G=\prod_{s}R_{s}^{n_{s}}/n_{s}!$). Since the $n_{1}$ single particles can be chosen in $N^{n_{1}}/n_{1}!$ ways, we get
\begin{eqnarray} \label{energy}
e^{\frac{-B}{kT}}=\sum_{n_{1}, n_{2}..}\prod_{s=2}\frac{R_{s}^{n_{s}}}{(n_{s}!)}y^{sn_{s}}p^{sn_{s}}p^{n_{1}}\frac{N^{n_{1}}}{n_{1}!} \nonumber\\
=e^{Np+\sum_{s=2}R_{s}(py)^{s}}
\end{eqnarray}
or
\begin{equation} \label{energy2}
B=-kT[Np+\sum_{s=2}R_{s}(py)^{s}]
\end{equation}
The average number of particles can be calculated from the partition function as $p(\partial$ln$q'/\partial p)$, giving us
\begin{equation} \label{number}
\bar{N}=Np+\sum_{s=2}sR_{s}(py)^{s}
\end{equation}
To calculate $R_{s}$, we first calculate $h_{s}$, the total number of ways where starting from a given atom, with successive steps taken to adjacent atoms, we return to the original atom after $s$ steps. Since we can start with any of the $N$ atoms, and a loop can begin at any of the $s$ atoms, we get $R_{s}=Nh_{s}/s$. Assuming a random walk, the probability per unit volume of being at a certain distance $r$ from the origin is given by $(2\pi d^{2}s/3)^{-3/2}e^{-3r^{2}/2sd^{2}}$. The probability of being back within an atomic volume $V/N$ of the origin ($r=0$) is then $(V/N)\Delta s^{-3/2}$, where $\Delta=(2\pi d^{2}/3)^{-3/2}$. If we further assume a lattice where each lattice site has $l$ neighbors, then the total number of walks after $s$ steps is $l^{s}$, but the number of walks that return to the origin are $h_{s}=(V/N)l^{s}\Delta s^{-3/2}$.
This gives us (with $\bar{N}=N$)
\begin{equation} \label{number2}
\frac{N}{V}=\frac{N}{V}p+\Delta \sum_{s=2} (lyp)^{s} s^{-3/2}
\end{equation}
All the arguments used to derive Eq.~\ref{number2} can be used for the colloidal case. Recall that for colloids, we have $y=e^{-d^2/2D\tau}$, while $l$ and $p$ remain conceptually the same as in the liquid helium case. The reason for the factor $n_{s}!$ in the denominator, from permutation statistics, is a bit more subtle. Following Frenkel \cite{Frenkel}, we can consider the loops to be similar to a distribution of polydisperse particles, that is, they can be placed within a bin of size$s$ containing all loops of that size. Permutations of loops within a certain bin size leave the macroscopic state of the system unaffected, and therefore, we should be able to divide individual terms in our partition function by $n_{s}!$. There is however one crucial difference between colloids and liquid helium atoms. The sum in Eq.~\ref{number2} begins at $s=2$, since permuting atoms to maintain symmetry requires at least two particles to be exchanged. For colloids, this is not the case, and we can look at displacements $d$ of single particles as well. We can therefore drop the first term on the right hand side of Eq.~\ref{number2}, and take the sum from $s=1$. Multiplying both sides of the equation by $(4/3)\pi a^{3}$, $a$ being the radius of the colloid, we then obtain for the colloidal case
\begin{eqnarray} \label{phig}
\phi=(4/3\pi) a^{3}\frac{N}{V}=(4/3\pi) a^{3}\Delta\sum_{s=1} (lyp)^{s} s^{-3/2}
\end{eqnarray}

To identify a glass transition, we note that the sum in the above equation converges when $lyp < 1$ and diverges when $lyp > 1$. At the glass transition where $\phi=\phi_{g}$, $lyp=1$ with loops spanning the entire volume.  This gives us $\phi_{g}=(4/3)\pi a^{3} \Delta \zeta(1.5)=0.17*2.612=0.44$ (since $d\approx 2a$ for nearly close packed spheres, $(4/3)\pi a^{3}\Delta=0.17$ and $\zeta$ is the Riemann-zeta function). If diffusive loops were to spontaneously appear in a colloidal system, our model predicts that we would see a glass transition, with these heterogeneous dynamic loops spanning the entire space at a volume fraction of 0.44. There are a few things to note in this derivation of $\phi_{g}$ for loops. First, the term $y$, and therefore the time $\tau$ does not enter in the calculation of $\phi_{g}$, making the transition independent of any time scale (albeit still a dynamic transition facilitated by dynamic heterogeneities). Second, we have not used any fit parameters to obtain the value of $\phi_{g}$, which is truly a remarkable result. Third, note that Eq.~\ref{phig} can be considered valid for the superfluid regime, that is, $lyp <1$ and $\phi < \phi_{g}$. For $\phi > \phi_{g}$, where large chains span the entire space, a different form of Eq.~\ref{phig} has to be considered. Finally, we should point out that our description of the glass transition assumes a `perfect gas' of loops, which may not be entirely accurate at high $\phi$, as we are neglecting interactions between loops in a crowded setting. However, we believe our model provides one of the first path integral based derivations of $\phi_{g}$ explicitly in terms of dynamic heterogeneities such as loops, and subsequent work can refine the above arguments further.

\subsection{Loop glass transition, fraction of particles participating}
In the previous section, we have assumed that \emph{all} particles participate in loops, i.e., they satisfy the constraint $\sum sn_{s}=N$. Here, we consider a new constraint $\sum sn_{s} = fN$ , where $f$ is the fraction of particles that are seen in loops. The reason for this is based on experiment. All experimental evidence to date in atomic, molecular and colloidal systems \cite{Glotzer98, Weeks00, Weeks02} suggest that dynamic heterogeneities (in these cases, strings) only occur among the fastest moving, most mobile particles, which typically are approximately $5\%$ of the total number of particles. In this context, it makes sense to calculate $\phi_{g}$ for loops where not all particles participate in them. We will assume however that the dominant contribution to the partition function will only be from the (most mobile) particles participating in loops. Therefore, if $Q$ is the partition function of all the particles, then $Q\simeq q$, where $q$ is the partition function from just the particles participating in loops. In other words, simply consider the partition function of the subset of particles that are highly mobile.

These assumptions will change the way our dynamic partition function is evaluated. In particular, we calculate $q'=p^{fN}Q\simeq p^{fN}q=e^{-B/k_{B}T}$. Equation~\ref{energy} can then be rewritten as
\begin{eqnarray} \label{final1}
e^{\frac{-B}{kT}}=\sum_{n_{1}, n_{2}..}\prod_{s=1}\frac{R_{s}^{n_{s}}}{(n_{s}!)}y^{sn_{s}}p^{sn_{s}}=e^{\sum_{s=1}R_{s}(py)^{s}}
\end{eqnarray}
or
\begin{equation} \label{final2}
B=-kT[\sum_{s=1}R_{s}(py)^{s}]
\end{equation}
The average number of particles can then be calculated simply by estimating $fN\simeq p (\partial$ln$q'/\partial p)$, giving us
\begin{equation} \label{final3}
fN\simeq\sum_{s=1}sR_{s}(py)^{s}
\end{equation}
Note that we could have derived Eq.~\ref{final3} simply by the transformation $N\rightarrow fN$ in Eq.~\ref{number} (and ignoring the first term). Now, if all particles were to participate in loops, we have $R_{s}=Nh_{s}/s$. If only a fraction of the particles \emph{can} be part of loops, then we have $R_{s}\approx fNh_{s}/s$. However, we still have $h_{s}=(V/N)l^{s}\Delta s^{-3/2}$. Note the absence of the factor $f$ in the denominator, since the volume per particle is always the same regardless of the fraction $f$. By this argument, $R_{s}$ scales as $f$ ($R_{s}\approx fVl^{s}\Delta s^{-5/2}$) for loops. Substituting the above into Eqn.~\ref{final3}, we obtain the same result for $\phi$ as in Eq.~\ref{phig}. Therefore, we obtain $\phi_{g}=0.44$ for all values of $f$. This is an interesting result, but perhaps not that surprising. Essentially, our arguments rests on the fact that the subset of particles that are mobile have behavior similar to the `bulk' case (that is, the distribution of loops scales as $f$ as does the number of mobile particles); however, the accessible volume available for this subset of particles is severely restricted by the presence of the other spectator particles. For this reason, we always get the glass transition at the same volume fraction regardless of the value of $f$. As a final point, we should note that for small values of $f$, interactions between loops become less important (as opposed to the case when all particles participated in loops), making our analysis somewhat more robust.

\subsection{Glass transition for strings and clusters, any number of particles participating}
While loops have been observed in simulations of molecular systems \cite{Glotzer98}, they are very few compared to the number of strings (or clusters) that are typically seen. We expect the same to be true for colloids as well. For a more realistic description of a colloidal glass transition, we will have to revisit Eq.~\ref{number} and evaluate the number of configurations that can be made in the system for strings and clusters, rather than loops.

It is immediately clear that loops place a highly restrictive condition on the number of possible configurations, since they have to return to the same location (within a volume $V/N$). In addition, circular symmetry for loops ensures that $R_{s}$ will contain a factor of $s$ in the denominator. On the other hand, counting the number of strings created by unrestricted random walks (analogous to a polymer chain) will always lead to divergent quantities. This is because for non-loop random walks $R_{s}\sim s^{-1.5}-s^{-1.8}$ (the upper limit being for self-avoiding walks) with no $s$ in the denominator, so that $\sum sR_{s}$ always diverges. This divergence can be addressed by noting that our strings exist in a supercooled matrix \cite{Langer2}. Due to this, the number of strings that exist will be greatly attenuated from the case of unrestricted random walks. However, it is difficult to come up with \textit{apriori} reasoning that gives us the form of $R_{s}$. Instead, we will need to approximate $R_{s}$ in Eq.~\ref{number} based on other work.

On examining Eq.~\ref{colloid} closely, the extension from loops to strings is relatively straightforward. For clusters, we note that the sum within the exponential in the integrand has to be modified to incorporate simultaneous motion of all particles within a cluster, summed over all cluster configurations. Then, it should be possible to use the same analysis for clusters as for strings and loops. All that remains is to sensibly approximate $R_{s}$ for clusters and strings. One such model we could turn to is Fisher's \emph{static} droplet model for condensation \cite{sator03,Fisher67}. This model assumes a set of non-interacting clusters of zero volume, in other words, a perfect gas of clusters (Feynman's model assumes a perfect gas of loops, a subset of the Fisher model). According to the model (for a more detailed description, see the Appendix), we have
\begin{equation} \label{Fisher2}
R_{s}=c_{0}Vs^{-\alpha} e^{\kappa' s-\gamma s^{\sigma}}
\end{equation}
where $\sigma$ characterizes the mean surface of clusters of size $s$ ($A\sim s^{\sigma}$, with $\sigma=2/3$ for a compact object, and $\sigma=1$ for a chain). $\kappa'$ and $\gamma$ are constants depending on the surface and volume energy and entropy of the clusters. We will use this static model and the associated form of $R_{s}$ for our dynamic heterogeneities.

Let us first consider the case of strings. For simplicity, let us also assume that all particles participate in the strings. For a string, $\sigma=1$, so that we can now rewrite Eq.~\ref{Fisher2} for dynamic strings as
\begin{equation} \label{Fisher}
R_{s}=c_{0}Vs^{-\alpha} e^{\kappa s}
\end{equation}
where $\kappa=\kappa'-\gamma$. In fact, Eq.~\ref{phig} (for loops) represents a special case of Fisher's equation for the critical density, where $\alpha = 2.5$, $c_{0}=\Delta$ and $e^{\kappa} \sim lyp$. It is important to note here that our dimensionless \emph{dynamic} parameter $y=e^{-d^{2}/D\tau}$ takes the place of static parameters in Fisher's model (this connection is made more explicit in the Appendix). Our equation for the colloidal volume fraction now becomes
\begin{equation} \label{prasad}
\phi=(4/3)\pi a^{3} c_{0} \sum_{s=1}(e^{\kappa}yp)^{s}s^{-\alpha+1}
\end{equation}
Similar to before, we assume the glass transition to occur when $e^{\kappa}yp = 1$. This gives us
\begin{equation} \label{prasad-2}
\phi_{g}=(4/3)\pi a^{3} c_{0} \sum_{s=1}s^{-\alpha}=(4/3)\pi a^{3} c_{0}\zeta(\alpha-1)
\end{equation}

To extrapolate the above expression for when only a fraction of particles participate in strings, we perform the transform $N\rightarrow fN$ and obtain
\begin{equation} \label{final4}
\phi_{g}\simeq(4/3)\pi a^{3} \frac{c_{0}}{f} \zeta(\alpha-1)
\end{equation}

For clusters, which are more compact than strings, we have $\sigma < 1$. Equation~\ref{prasad} can then be rewritten as
\begin{equation} \label{prasad-cluster}
\phi=(4/3)\pi a^{3} c_{0} \sum_{s=1}(e^{\kappa}yp)^{s}e^{-\gamma s^{\sigma}}s^{-\alpha+1}
\end{equation}
Here, to identify the glass transition, we need to satisfy both conditions $e^{\kappa}yp = 1$ and $e^{-\gamma}=1$. Then, our expression for $\phi_g$ will be the same as for the case of strings, for all values of $f$ (that is, the same as Eqns.~\ref{prasad-2} and ~\ref{final4}). Note that similar to the case of loops, the value of $c_{0}$ will adjust itself for different values of $f$ so that $c_{0}/f$ is always a constant.

We are now ready to connect our theoretical predictions with experiment. In order to do so, we will need to use the values of $c_{0}$ and $\alpha$ for the case of clusters from experiment (we have already noted that the value of $f$ should not matter). Unfortunately, the limited data to date~\cite{Weeks00} only provides the power law exponent $\alpha$ ($\alpha=2.2$) and not the value of $c_{0}$, with the distribution of cluster sizes being given in a normalized form. Because of this, an assumption will have to be made to approximate this parameter, which is that the value of $c_{0}$ does not change substantially when $\alpha$ is changed slightly. Recall that for loops with $\alpha=2.5$, $(4/3)\pi a^3 c_{0}/f=0.17$ for all values of $f$. It then follows from our assumption that $\phi_{g}=0.17\zeta(\alpha-1)$. Fitting for the known value of $\phi_{g}=0.58$, we obtain $\alpha=2.356$. Alternately, we could fit for $c_{0}$, assuming that $\alpha=2.2$, giving us $(4/3)\pi a^3 c_{0}/f=0.58/\zeta(1.2)=0.104$ (to be compared to the value of 0.17 for diffusive loops). Considering all the approximations that have gone into our model, we believe our theory makes reasonable predictions, but can probably be improved with better alignment with experiments. For future work, however, we would certainly recommend that experimental researchers provide the raw distribution of clusters so that the number of configurations can be exactly determined.

In the subsequent section, we will use the forms of $\phi_{g}$ determined above to describe the glass transition under confinement, which may provide further insights into which form of $R_{s}$ is more accurate.

\subsection{Glass transition under confinement}
An analysis of colloidal heterogeneities under confinement requires substantially more depth than we are able to provide in this Article; however, we will sketch the beginnings of an approach that may provide some insights into the matter. Specifically, we will try to calculate how reducing the number of available configurations available for loops/strings/clusters by confining colloids in a box (say of dimension $L$) will lower $\phi_{g}$. To begin with, let us consider a system with `bulk' volume fraction $\phi$, and for simplicity, only consider the case of loops. If the system were unconstrained, then according to Eq.~\ref{phig}, $\phi=0.17\sum_{s=1} (lyp)^{s} s^{-3/2}$. In other words, the distribution of loops will not be an exact power law, and we will have $lyp < 1$. For instance, if the volume fraction $\phi=0.36$, $lyp\approx 0.98$. Under confinement, when the box size $L$ is made smaller and smaller, the dynamics of the loops will slow down and we will start approaching the glass transition. In experiment~\cite{Zhu12}, one of the ways to determine the critical length scale at which this occurs is by noting where the $\alpha$-relaxation time $\tau_{\alpha}$ becomes the same as the relaxation time at $\phi=0.58$, $\tau_{\alpha,bulk}$. For our model, an equivalent assumption is when the loop distribution under confinement mimics that at $\phi_{g}$. In other words, $lyp$ will adjust itself so that $lyp=1$ and $R_{s}$ will follow a power-law distribution. Since our clusters cannot be larger than the box dimension, we will have a natural cut-off where $s< L/a$. Then, we have for a glass former under confinement
\begin{equation} \label{confinement}
\phi_{g,c} \approx 0.17 \sum_{s=1}^{L/a}s^{-3/2}=0.17H_{M}^{1.5}
\end{equation}
where $M=L/a$ is a dimensionless parameter, and $H$ is the generalized harmonic number. To illustrate the above better, consider our specific example where $\phi = 0.36$. For this volume fraction to demonstrate a glass transition, we need to have $H_{M}^{1.5}=0.36/0.17$, or $M\approx 16$ by fitting for $M$.

The above analysis can be extended to dynamic clusters, in order to make connections with experiment. There are two possibilities for clusters: 1) For $\alpha=2.356$, $\phi_{g,c}=0.17H_M^{1.356}$ 2) For $\alpha=2.2$, $\phi_{g,c}=0.104H_M^{1.2}$. Again, for the specific case of $\phi=0.36$, a glass transition will occur (for the two given possibilities) when 1) $M=8$ or 2) $M=70$. Indeed, while the bulk $\phi_{g}$ can be obtained by infinitely many combinations of $c_{0}$ and $\alpha$, $\phi_{g,c}$ is very sensitive to the choice of these two parameters.
\begin{figure}
\includegraphics[scale=0.6]{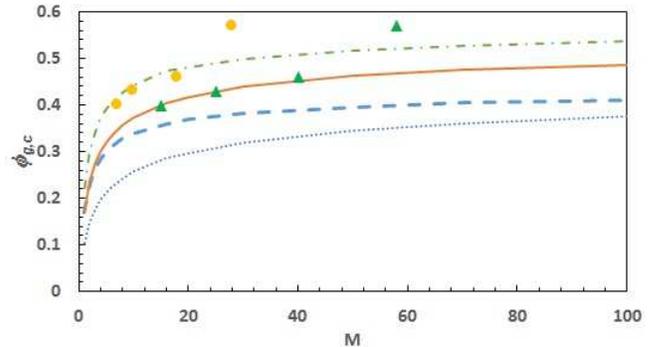}
\caption{\label{foo} $\phi_{g,c}$ plotted against the normalized box dimension $M$ for various cases. Symbols are: dashed and dash-dot lines, loops under confinement ($\alpha=2.5$); solid and dotted lines, clusters under confinement ($\alpha=2.356$ and $\alpha=2.2$ respectively); triangles, experimental data from~\cite{Zhu11}; circles, experimental data from~\cite{Zhu12}.}
\end{figure}
We emphasize this point in Fig.~\ref{foo}, which plots the normalized box dimension $M$ against $\phi_{g,c}$ for the two possibilities for clusters (solid and dotted lines). In addition, the dashed line shows how diffusive loops behave under confinement. We have also included a curve for loops (dash-dot line), where we have force-fitted $\phi_{g}=0.58$, which gives $\phi_{g,c}=0.22H_M^{1.5}$. Finally, from references~\cite{Zhu11, Zhu12}, we plot experimental data for $\phi_{g,c}$ that is obtained in two different ways. The circles are the normalized critical length scale as described before ($\tau_{\alpha}\sim\tau_{\alpha,bulk}$), for four volume fractions $\phi= 0.4$, $0.43$, $0.46$ and $0.57$ respectively. The triangles are a confinement length scale determined from mean square displacements, four-point susceptibility and $\alpha_{2}$ measurements where deviations from bulk behavior occur, at the same four volume fractions.

As mentioned before, the value of $\phi_{g,c}$ is very sensitive to the choice of $c_{0}$ and $\alpha$. From the graph, it is clear that while three of the theoretical curves asymptote to $\phi_{g} =0.58$ as $M\rightarrow\infty$, they deviate from each other at intermediate length scales. Since there is such a wide spread in the different curves, it is not surprising that we can find combinations of $c_{0}$ and $\alpha$ that fit the experimental data. Indeed, we find that the experimental data can be reasonably fit for $\phi<0.58$ by either an expression for loops or for clusters. This further emphasizes the point that closer connection with experiment will be critical for further refinement of our theory. Still, our theory provides a simple framework for thinking about the colloidal glass transition under confinement, namely, the restriction of configurations.
\section{Conclusions}
The main result of our approach is the demonstration that a glass transition \textbf{\emph{can}} arise from the presence of dynamic heterogeneities in the form of loops/strings/clusters, even if only a fraction of particles are participants. Indeed, our model looks \textbf{\emph{only}} at the configurations of the heterogeneities and ignores contributions for particles that are spectators. It is essential that these heterogeneities follow a power-law distribution; with this assumption we can reproduce the classic colloidal glass transition volume fraction, $\phi_{g}=0.58$, with sensible choices for the form of this distribution.

Compared to other, more detailed theories~\cite{Parisi} of the colloidal glass transition, our `path-integral' approach may be considered quite simplistic. For instance, we are assuming effective free-particle motion within our dynamic chains, disregarding the effects of any interaction potential. The motion within the heterogeneities is also considered to be the only important motion, and that the diffusive motion of other particles does not matter (this is easier to justify for large rather than small $f$). Our approach also does not shed any light as to why strings or clusters of particles may spontaneously appear near a colloidal glass transition. The use of Fisher's model for our dynamic system, while a reasonable assumption, may require stronger justification as well. In spite of these caveats, the approach we have followed may yet provide some guidance towards relating dynamic heterogeneities to the colloidal glass transition. Since our model considers only the dynamic heterogeneities and not the other spectator particles, calculating the effects of confinement also becomes relatively straightforward.

Future work will involve closer connection to experiment to improve the fits of our model. We will also attempt to make our model more rigorous; for instance, we will incorporate the motions of all particles and not just those in the heterogeneities. We will also attempt to include the effects of an interaction potential as a perturbation to the free-particle case to encompass more glass-forming situations. Finally, it should be possible to use our expression for the partition function to generate quantities involving correlations, such as $\chi_{4}$, to connect our theory with the vast body of literature , both theoretical and experimental, that already exists for these systems.
\appendix*
\section{Fisher model} \label{append}
The Fisher model assumes a perfect gas of clusters, and in this context, the grand partition function of the system can be written as
\begin{equation} \label{app-1}
\Xi(z,T,V)=e^{\sum_{s=1}^{\infty}q_{s}z^{s}}
\end{equation}
where $q_{s}(T,V)$ is the partition function of a cluster of size $s$, and $z=e^{\beta \mu}$ is the fugacity of the particles ($\mu$ being the chemical potential). A detailed derivation of Eq.~\ref{append} can be found in the Appendix of reference~\cite{sator03}. From this partition function, thermodynamic quantities such as the pressure, specific heat and (important for our purposes) density can be calculated. It then remains to find suitable expressions for $q_{s}(T,V)$ based on properties of the clusters. Typically, this is done by noting that $q_{s}(T,V)=e^{-\beta F_{s}(T,V)}$, where $F_{s}$ is the free energy of a cluster of size $s$.

Very briefly, the internal energy and entropy of a cluster of size $s$ can be written as a sum of surface and volume terms~\cite{sator03}
\begin{eqnarray} \label{app-2}
U_{s}=-u_{v}s+u_{a}A_{s}\nonumber\\
S_{s}=s_{v}s+s_{a}A_{s}
\end{eqnarray}
where $u_{v}$, $s_{v}$, $u_{a}$ and $s_{a}$ are volume energy and entropy, and surface energy and entropy per particle respectively, and $A_{s}$ is the surface area of a cluster. A parameter $\sigma$ can be used to characterize the mean surface area of the clusters, so that $A_{s}=a_{0}s^{\sigma}$ with $0<\sigma<1$. Finally, a corrective logarithmic term is added to the free energy (in a somewhat ad-hoc fashion), so that the free energy can be written as
\begin{equation} \label{app-3}
-\beta F_{s} = \beta (u_{v}+s_{v}T)s-\beta a_{0}(u_{a}-s_{a}T)s^{\sigma}-\alpha \text{ln} (s) + \text{ln} (c_{0} V)
\end{equation}
with the term proportional to ln $V$ resulting from the integration over the center of mass of the cluster, $c_{0}$ being a constant. With this, the partition function of a cluster is given by
\begin{eqnarray} \label{app-5}
&q_{s}(T,V)=e^{-\beta F_{s}(T,V)} \nonumber\\
&=c_{0}Vs^{-\alpha}e^{\beta(u_{v}+s_{v}T)s-\beta a_{o}(u_{a}-s_{a}T)s^{\sigma}}
\end{eqnarray}

This derivation can be used to connect Fisher's model to Feynman's model, and our dynamic strings and clusters as well. To do this, note that the free energy of a particle moving in liquid helium is simply its kinetic energy; in other words, $F\sim(1/2)m'v^{2}$. For a distance $d$ travelled in a time $t$, we have $v=d/t=d/(\beta\hbar)$ or $F=(1/2)m'd^{2}/(\beta\hbar)^2$. For $s$ particles, we then have $F_{s}=sF$ or $q_{s}=e^{-\beta F_{s}}\sim e^{-(m'd^{2}/2\beta \hbar^{2})s}$, or $q_{s}\sim y^{s}$, where $y$ is as before in Feynman's derivation. In this way, we have connected Fisher's static model to Feynman's model, and by extension, to our dynamic loops and strings. More explicitly, comparing the forms of the partition function in Eq.~\ref{app-1} and Eq.~\ref{final1}, we can say $q_{s}\sim R_{s}(py)^{s}$, with the form of $q_{s}$ in Eq.~\ref{app-5} finally giving us the expression for $R_{s}$ in Eq.~\ref{Fisher}.

For dynamic clusters, note that the volume term is exactly the same, in other words, the free energy from the motion (or diffusion) of $s$ particles within the cluster will be proportional to $s$. The area term is more difficult to justify, as there is no easy analog for the motion of a cluster that involves its area (except perhaps viscous drag). However, we will leave the term in for the dynamic case to ensure some flexibility in our model to incorporate solvent effects (say), while noting that its presence makes no difference to the location of the glass transition since we choose $e^{-\gamma}=1$. With this in mind, it should be feasible to use Eq.~\ref{Fisher2} for dynamic clusters.

%

\end{document}